\documentclass[12pt]{iopart}

\usepackage{epsfig}
\usepackage{graphicx}
\usepackage{dcolumn}
\usepackage{bm}
\usepackage{graphics}

\begin{document}

\title{Effect of on-site Coulomb interaction (\textit{U}) on the electronic and magnetic properties of Fe$_{2}$MnSi, Fe$_{2}$MnAl and Co$_{2}$MnGe}
\author{Sonu Sharma and Sudhir K. Pandey}
\address{School of Engineering, Indian Institute of Technology Mandi, Kamand - 175005, India}
\address{Electronic mail: sonusharma@iitmandi.ac.in}

\date{\today}

\begin{abstract}
The electronic band structures, density of states plots and magnetic moments of Fe$_{2 }$MnSi, Fe$_{2 }$MnAl, and Co$_{2 }$MnGe are studied by using the first principles calculations. The FM solutions using LSDA without \textit{U} show the presence of half-metallic ferromagnetic (HFM) ground state in Fe$_{2 }$MnSi, whereas the ground state of Fe$_{2 }$MnAl is found to be metallic. In both compounds the maximum contribution to the total magnetic moment is from the Mn atom, while the Fe atom contributes very less. The electronic structures and magnetic moments of Fe-based compounds affected significantly by \textit{U}, whereas its effect is very less on Co$_{2}$MnGe. The magnetic moment of Fe atom in Fe$_{2 }$MnSi (Fe$_{2 }$MnAl), increased by $\sim$ 70 \% ($\sim$ 75 \%) and in Mn atom it decreases by $\sim$ 50 \% ($\sim$ 70 \%) when the value of \textit{U} is increased from 1 to 5 eV. The Hund's like exchange interactions are increasing in Fe atom while decreasing in Mn atom with increase in \textit{U}. The Fe and Mn moments are ferromagnetically coupled in Fe$_{2 }$MnSi for all values of \textit{U}, whereas in Fe$_{2 }$MnAl they coupled antiferromagnetically below \textit{U} = 2 eV and ferromagnetically above it. Above \textit{U} = 2 eV the metallic ground state of Fe$_{2 }$MnAl changes to semiconducting ground state and the ferromagnetic coupling between Fe and Mn atoms appears to be responsible for this.
\end{abstract}


\maketitle

\section{Introduction}
After the discovery of half-metallic ferromagnetism in half-Heusler alloy NiMnSb by de Groot et al.\cite{groot} many of the compounds were found to be half-metallic in different experimental \cite{hanseen,kirillova} and theoretical studies\cite{igalanakis,kubler,ishida,fujii,picozzi,barth,candan}. The half-metallic ferromagnets (HMFs) have band gap at the Fermi level (\textit{E$_{F}$}) in one spin channel while the other spin channel is strongly metallic. These materials show a complete spin polarization of the conduction electrons at the \textit{\textit{E$_{F}$}}. Because of their exceptional band structures at the \textit{E$_{F}$}, these materials are of great interest from theoretical and applications point of view. Co containing full-Heusler alloys were firstly proposed by Ishida et al.\cite{ishida} and were synthesised by Webster\cite{webster}. Fe$_{2 }$MnZ type Heusler alloys have also been proposed to show half-metallic ferromagnetism by Fujii et al.\cite{fujii}. In the family of Fe based Heusler alloys only Fe$_{2 }$MnSi\cite{fujii} and Fe$_{2 }$CrZ (Z = Si, Ge, Sn) \cite{sishida} were predicted to be half-metallic ferromagnts theoretically. Fe$_{2 }$MnSi is a ferromagnetic material with a Curie temperature of 214 K and transforms at 69 K to the low temperature phase with smaller spontaneous magnetization \cite{ziebeck}. The total magnetic moment of Fe$_{2 }$MnSi obtained by magnetization measurements is 2.1 $\mu_{B}$/f.u. (f.u.$\equiv$ formula unit) at 4.2 K \cite{kawakami} which is smaller than the calculated moment of 3.0 $\mu_{B}$/f.u. \cite{sfujii,bhamad}. Fe$_{2 }$MnAl on the other hand is not a half-metallic as there exist slight density of states at the \textit{E$_{F}$} for minority spin channel \cite{fujii}. 

It is well known that the local spin density approximation (LSDA) and the generalized gradient approximation (GGA) schemes for the exchange-correlation potential are not sufficient to describe the electronic structure and magnetism of some Heusler alloys like Co$_{2 }$FeSi\cite{kandpal,wurmehl}. Such strongly correlated systems which contain atoms with open \textit{d} or \textit{f} shells, can be treated by adding on-site Coulomb interaction (\textit{U}) term as modification to LSDA i.e. by using LSDA + \textit{U} approach \cite{czyzyk}. The LSDA + \textit{U} method accounts an orbital-dependent on-site electron-electron Coulomb interaction which is not included in the pure LDA or GGA. Rai et al.\cite{rai, dprai} have studied some Co based Heusler alloys and reported the increase in band gap, hybridization of \textit{d}-\textit{d} orbitals as well as \textit{d}-\textit{p} orbitals when on-site Coulomb interaction is added to LSDA approach. They have also concluded that some Co-based Heusler alloys show half-metallic character when are treated with LSDA + \textit{U}. It is well known that the systems in which there exist the density of states (DOS) at the Fermi energy for one spin, the on-site Coulomb interaction (\textit{U}) may bring a drastic change into their electronic and magnetic properties \cite{ssishida}. As the electron-electron correlation plays an important role in the Heusler compounds so it can be expected that it will also affect the Fe-based Heusler alloys.   

In the present work we have employed the full-potential linearized augmented-plane wave methods to compute the electronic and magnetic properties of three full-Heusler alloys viz. Fe$_{2 }$MnSi, Fe$_{2 }$MnAl, and Co$_{2 }$MnGe. The electronic structures in the FM solutions without \textit{U} show that Fe$_{2 }$MnSi is half-metallic compound whereas Fe$_{2 }$MnAl is metallic. The large effect of \textit{U} is observed on the electronic structures and magnetic moments of Fe-based compounds whereas on Co$_{2 }$MnGe compound its effect is negligibly small. The magnetic moment of Fe atom increases with \textit{U} and in the Mn atom it decreases with \textit{U}. The magnetic moments of Fe and Mn atoms are found to be coupled ferromagnetically in Fe$_{2 }$MnSi for all values of \textit{U}, whereas in Fe$_{2 }$MnAl they coupled antiferromagnetically below \textit{U} = 2 eV and ferromagnetically above it. The ground state of Fe-based compounds remains half-metallic for all values of \textit{U}, whereas it changes from metallic to semiconductor in Fe$_{2 }$MnAl after \textit{U} = 2 eV. Also the ferromagnetic coupling between Fe and Mn atom is found to be responsible for the presence of semiconducting ground state in this compound. 

\section{Computational details and Crystal structure}
The electronic and magnetic properties of Fe$_{2}$MnSi, Fe$_{2}$MnAl and Co$_{2}$MnGe were calculated by using the full-potential linearized augmented plane-wave (FP-LAPW) method within the density functional theory (DFT) implemented in WIEN2k code\cite{blaha}. The local spin density approximation (LSDA) of Perdew and Wang \cite{perdew} was employed for exchange-correlation energy of electrons. The effect of on-site Coulomb interaction (\textit{U}) under LSDA +\textit{U} \cite{czyzyk} formulation of the DFT was also considered in the calculations. The around-the-mean-field (AMF) version of the LSDA+\textit{U} \cite{czyzyk} method was employed to account for the “double-counting” correction terms in the energy functional. The effective Coulomb-exchange interaction \textit{U}$_{eff}$ is given by (\textit{U } - \textit{J}), where \textit{U} and \textit{J} are the Coulomb and exchange parameter. The value of \textit{U} was varied from 1 to 5 eV and \textit{J} was kept fixed to 0 eV, therefor \textit{U}$_{eff}$ was equal to \textit{U} in our calculations. The values of muffin-tin radii used in the calculations were 2.2 Bohr for Fe, Mn, Si and Al atoms. The maximum \textit{l} value (\textit{l$_{max}$}) for the expansion of wave function in spherical harmonics inside the atomic spheres was equal to 10. For convergence of energy eigenvalues the wave function in the interstitial regions were expanded in plane waves with cutoff $R_{mt}K_{max}$ = 8, where $R_{mt}$ is the smallest atomic sphere radius and $K_{max}$ is largest k vector in the plane wave expansion. The electronic and magnetic properties of these compounds were calculated by using the experimental lattice parameters. The self consistent iteration was repeated until calculated total energy/cell and charge/cell of the systems converge to less than 10$^{-4}$ Ry and 0.001e, respectively.

All these compounds belong to the family of full Heusler alloys and crystallizes in $L2_{1}$ crystal structure with space group $Fm$-$3m$. These compounds have composition X$_{2 }$YZ, where X and Y are transition metals and Z is the main groups element. X atoms (Fe and Co) are placed at Wyckoff position 8c (1/4, 1/4, 1/4). Y atoms (Mn) and Z atoms (Si, Al and Ge) are located at Wyckoff position 4a (0, 0, 0) and 4b (1/2, 1/2, 1/2), respectively \cite{brown,hkandpal}.

\section{Results and discussions}
\subsection{Paramagnetic Phase}

The dispersion curves of Fe$_{2}$MnSi and Fe$_{2}$MnAl along the high symmetry directions of the first Brillouin zone are presented in Fig. 1(a and b). In the dispersion curve of Fe$_{2}$MnSi, the bands labeled by 1, 2 and 3 are lying above the \textit{E$_{F}$}, bands 4, 5, 6 and 7 are crossing it at 14 different k-points and band 8 is lying below it. Around W-point, bands 4-7 are concentrated in the energy range of about -0.2 to 0.2 eV. The total energy of the system can be minimized if there will be shifting of these bands and this shifting may lead to the FM ground state for this compound. From Fig. 1(b) it is clear that in Fe$_{2}$MnAl the bands labeled by 1-5 are lying above the \textit{E$_{F}$}, bands 6 and 7 which are crossing the \textit{E$_{F}$} at 10 different k-points and band 8 is lying just below it. Bands 4-7 are concentrated around the W-point in the energy range of about -0.2 to 0.2 eV. Therefore here also one can expect that shifting of these bands will minimize the total energy of the system which may lead to the ferromagnetic ground state in the compounds. These results are consistent with our earlier reported results for the paramagnetic phase of Co$_{2}$MnGe\cite{sonu}. 

The total and partial density of states plots of Fe$_{2}$MnSi and Fe$_{2}$MnAl are presented in Fig. 2 and Fig. 3, respectively. From total density of states plots (TDOS) (Fig. 2(a) and 3(a)) it is clear that there is very large density of states of about 6 states/eV/f.u. (f.u.$\equiv$ formula unit) at \textit{E$_{F}$} for both the spins. According to the Stoner theory the large value of TDOS may be considered as the indication of the ferromagnetic ground states in the compounds\cite{skpandey,sonu}. The antibonding bands are extended upto 0.4 eV and 0.3 eV below the \textit{E$_{F}$} for Fe$_{2}$MnSi and Fe$_{2}$MnAl, respectively. As per Stoner theory the total energy of the systems will be minimized if there is a shifting in spin-up and spin-down bands by $\sim$0.4 eV and 0.3 eV below and above the \textit{E$_{F}$} for Fe$_{2}$MnSi and Fe$_{2}$MnAl, respectively. This may be responsible for the half-metallic FM ground state in the compounds as observed in the case of Co$_{2}$MnGe\cite{sonu}. Also the total energies of FM phase of Fe$_{2}$MnSi and Fe$_{2}$MnAl are about 0.77 eV and 0.54 eV less than PM phase, which further confirm that both compounds should have FM ground states. 
The partial density of states (PDOS) plots for Fe, Mn and Si atoms of Fe$_{2}$MnSi are presented in Fig. 2(b-d). From Fig. 2(b) it is evident that the PDOS of Fe atom at \textit{E$_{F}$} is mainly contributed by \textit{\textit{t$_{2g}$}} and \textit{e$_{g}$} states with contribution of about 0.3 states/eV/atom and about 2.0 states/eV/atom, respectively for both spin channels. The PDOS of Mn atom is shown in Fig. 2(c) and in both spin channels the contribution from \textit{t$_{2g}$} and \textit{e$_{g}$} is $\sim$1 states/eV/atom and $\sim$1.6 states/eV/atom, respectively. The PDOS of Si atom (Fig. 2(d)) show that the occupancy of 3\textit{s} and 3\textit{p} orbitals at \textit{E$_{F}$} is very small, which can be neglected. The PDOS plots for Fe, Mn and Al atoms of Fe$_{2}$MnAl are shown in Fig. 3(b-d). It is clear from Fig. 3(b) that the contribution of \textit{t$_{2g}$} is $\sim$0.3 states/eV/atom and \textit{e$_{g}$} is $\sim$1 states/eV/atom at \textit{E$_{F}$} for both spin channels. In the PDOS of Mn (Fig. 3(c)) the occupancy of \textit{t$_{2g}$} at \textit{E$_{F}$} is $\sim$0.7 states/eV/atom and \textit{\textit{e$_{g}$}} at \textit{E$_{F}$} is $\sim$2.0 states/eV/atom. From PDOS plot of Al atom (Fig. 3(d)) it is evident that there is negligibly small contribution from 3\textit{s} and 3\textit{p} orbitals at \textit{E$_{F}$}. 
It is also clear from these figures that \textit{t$_{2g}$} and \textit{e$_{g}$} have main contribution to the total DOS at \textit{E$_{F}$} for Fe$_{2}$MnSi and Fe$_{2}$MnAl. The \textit{e$_{g}$} state has the largest contribution to the TDOS of both compounds at \textit{E$_{F}$}.

\subsection{Ferromagnetic Phase}
The spin resolved dispersion curves of Fe$_{2}$MnSi and Fe$_{2}$MnAl along the high symmetry directions of the first Brillouin zone are shown in Fig. 4. From the dispersion curve shown in Fig. 4(a) it is clear that Fe$_{2}$MnSi is metallic for spin-up channel. Bands labeled by 1 and 2 are lying just above the \textit{E$_{F}$}, bands 3 and 4 are crossing the \textit{E$_{F}$} at 7 different k-points and bands 5-7 are lying just below it.  Fig. 4(b) shows that this compound behave as semiconductor for down spin channel and there exist an indirect gap of about 0.44 eV from $\Gamma$ to X-direction. The computed value of indirect band gap using LSDA is less than the reported value \cite{hongzhi}. This compound is found to be half-metallic similar to the previous studies \cite{fujii,hamad}. Bands labeled by 1-6 have shifted into the conduction band while bands 7 and 8 have shifted into the valence band and thus there is presence of ferromagnetic ground state in this compound as stated earlier. From the dispersion curves shown in Fig. 4(c and d), it is very clear that Fe$_{2}$MnAl is metallic in nature. In the spin-up channels bands 1-4 are lying above the \textit{E$_{F}$} and bands 5-8 are crossing it at 8 different k-points. In down spin channel bands 1-7 are lying above \textit{E$_{F}$} and only one band, labelled by 8 is crossing it at two different k-points. Thus there exist a slight density of states at \textit{E$_{F}$} in the spin-dn channel and making this compound metallic in nature. There is almost flat conduction band along $\Gamma$ to X direction in the down spin channel of both compounds and this can be responsible for large value of effective masses of these compounds. 

Total and partial DOS plots of Fe$_{2}$MnSi for FM solution are shown in Fig. 5. From Fig. 5(a), it is clear that spin-up channel is occupied at \textit{E$_{F}$} with occupancy of about 5 states/eV/f.u. whereas spin-down channel is unoccupied. Thus this compound behave as metal for majority spin states and semiconductor for minority spin states. After comparing Fig. 2(a) and 5(a), one can conclude that TDOS shifts towards lower energy in spin-up channel whereas, in spin-down channel it shifts towards higher energy. Because of this shift there is creation of band gap in the minority spin channel as is observed in Co$_{2}$MnGe. This band shift appears to be responsible for the presence of FM ground state in this compound. However, in this compound the shift in TDOS is very small in comparison to rigid shift observed in spin-up channel of Co$_{2}$MnGe\cite{sonu}. Also the value of TDOS at \textit{E$_{F}$} is very large in comparison to TDOS of Co$_{2}$MnGe studied earlier and one can expect very large effect of on-site Coulomb interaction (\textit{U}) on such systems. From the PDOS of Fe atom (Fig. 5(b)) it is clear that \textit{t$_{2g}$} ($\sim$0.5 states/eV/atom) and \textit{e$_{g}$} ($\sim$1.4 states/eV/atom) states have main contribution at \textit{E$_{F}$} for spin-up channel while the minority spin channel is empty. Near the \textit{E$_{F}$} in spin-dn channel, the valence band maximum has main contribution from \textit{t$_{2g}$} and conduction band minimum is contributed by \textit{e$_{g}$} states. The PDOS of Mn atom is presented in Fig. 5(c) and it is clear from this plot that \textit{t$_{2g}$} bands have occupation of $\sim$1.0 states/eV/atom with negligibly small contribution from \textit{e$_{g}$} band at \textit{E$_{F}$} for spin-up channel. The \textit{e$_{g}$} band of Mn atom has shifted towards lower energy with no contribution at the \textit{E$_{F}$}. The PDOS of Si atom (Fig. 5(d)) shows negligibly small contribution from 3\textit{s} and 3\textit{p} orbitals. 

The total and partial density of states plots of spin-up and spin-down channels for Fe$_{2}$MnAl are shown in Fig. 6. The spin-up channel of Fig. 6(a) shows that total density of states at \textit{E$_{F}$} is $\sim$1 states/eV/f.u. and spin-down channel show very small ($\sim$0.1 states/eV/f.u.) density of states at \textit{E$_{F}$}. On comparing Fig. 3(a) and 6(a) one can find that there is no rigid shifting of bands rather there is splitting of bands at \textit{E$_{F}$}, which we have not observed in Co$_{2}$MnGe and Fe$_{2}$MnSi. The PDOS of Fe, Mn and Al atoms for both spin channels are shown in Fig. 6(b-d). From Figs. 3(b) and 6(b) it is clear that there is no rigid shift in band as per Stoner theory as is observed in Co$_{2}$MnGe\cite{sonu}. The \textit{e$_{g}$} states split in such a way that there exists a minimum at \textit{E$_{F}$} with contribution of $\sim$0.3 states/eV/atom and 0.1 states/eV/atom, from \textit{t$_{2g}$} and \textit{e$_{g}$} bands, respectively. This minimum is responsible for the existence of the pseudo gap. In the minority spin channel the valence band maximum has mainly \textit{t$_{2g}$} character and conduction band minimum has \textit{e$_{g}$} character. This compound show metallic nature for down spin also because of very small contribution from \textit{t$_{2g}$} bands at \textit{E$_{F}$}. In Fig. 6(c) the PDOS for Mn \textit{t$_{2g}$} and \textit{e$_{g}$} bands are shown. The spin-up channel of PDOS of Mn atom is contributed by \textit{t$_{2g}$} state ($\sim$0.2 states/eV/atom) with negligibly small contribution from \textit{e$_{g}$} state at \textit{E$_{F}$}. It is evident from Fig. 6(d) that in PDOS of Al atom there is negligibly small contribution of 3\textit{s} and 3\textit{p} orbitals.

The total magnetic moment per formula unit for Fe$_{2}$MnSi is 3.0 $\mu_{B}$ with contribution from Fe, Mn, Si and interstitial region is 0.22, 2.52, -0.01 and 0.042 $\mu_{B}$, respectively. The similar value of total magnetic moment is also predicted theoretically in \cite{galanakis,bhamad}, whereas the experimental results \cite{ueda,lhongzhi} gave a saturation magnetic moment less than this value. This may be due to the reason that it is difficult to obtain the pure phases experimentally. The Mn atom coupled ferromagnetically with Fe atom as is found by Galanakis et al.\cite{galanakis}. The magnetic moment of Si is very small and it is coupled antiferromagnetically with Fe and Mn atoms. The total magnetic moment per formula unit for Fe$_{2}$MnAl is 2.0 $\mu_{B}$ and contribution from Fe, Mn, Al and interstitial region is -0.23, 2.44, -0.008 and 0.05 $\mu_{B}$, respectively. The calculated value of the total magnetic moment matches with earlier reported value \cite{galanakis}. The Mn atom coupled antiferromagnetically with Fe atom in this compound. By using the full-potential screened Korringa-Kohn-Rostoker (FSKKR) Green’s function method in conjunction with the local spin density approximation Galanakis et al.\cite{galanakis} have also found similar results. But Fujii et al. \cite{fujii} have observed ferromagnetic coupling between Fe and Mn atoms which is contradictory to our results. The magnetic moment of Al is very small but it is coupled ferromagnetically with Fe atom and antiferromagnetically with Mn atom. The Mn atom carries the largest magnetic moment in all these compounds, while the Fe atom of Fe$_{2}$MnSi and Fe$_{2}$MnAl carry the modest magnetic moment which is in agreement with the earlier reported results \cite{plogmann}. 

\subsection{Effect of on-site Coulomb interaction (\textit{U})}

From the study of electronic structure of Fe$_{2}$MnSi it is clear that for spin-up channel there is very large density of states at \textit{E$_{F}$} and spin-down channel is empty. The electronic structure of Fe$_{2}$MnAl show comparatively small value of TDOS from 3\textit{d} bands at \textit{E$_{F}$} for both spin channels. It is well known that 3\textit{d} bands are less dispersive therefore on-site Coulomb interaction can have very large effect on the electronic and magnetic properties of such compounds. The LSDA+\textit{U} method may bring a drastic change on the magnetic properties and electronic properties of these compounds. So we have study the effect of \textit{U} varying from 1 to 5 eV on the magnetic moment and electronic structure of these compounds along with previously studied Co$_{2 }$MnGe compound \cite{sonu}. 
 
Firstly, we discuss the effect of \textit{U} on the magnetic moments of all these compounds. The total number of 3\textit{d} electrons in both spin channels and local magnetic moments in the presence of \textit{U} are presented in Table 1 for Fe$_{2}$MnSi, Fe$_{2}$MnAl and Co$_{2}$MnGe. It is clear from Table that in Fe$_{2}$MnSi, the number of 3\textit{d} electrons of Fe atom in up spin channel increases while that of down spin channel decreases with \textit{U}. However the total number of electrons remains fixed to the value $\sim$6.1. The local magnetic moment of Fe atom is found to increase with increasing the value of \textit{U}. The magnetic moment increases from 0.29 $\mu_{B}$ at \textit{U} = 1 eV to 0.87 $\mu_{B}$ at \textit{U} = 5 eV. In Mn atom, the total number of 3\textit{d} electrons in spin-up channel decreases and spin-dn channel increases with \textit{U} in such a way that total number of 3\textit{d} electrons remains fixed to the value $\sim$4.9. Also the magnetic moment of Mn atom decreases from 2.41 $\mu_{B}$ at \textit{U} = 1 eV to 1.29 $\mu_{B}$ at \textit{U} = 5 eV. The above results suggest that the Hund's like exchange interactions between Fe 3\textit{d} electrons are increasing and that of Mn 3\textit{d} electrons decreasing with increase in \textit{U}.
  
In Fe$_{2}$MnAl, also the number of electrons of Fe (Mn) atom in up spin channel increases (decreases) while that of down spin channel decrease (increases) with \textit{U}. The value of local magnetic moment of Fe atom also goes on increasing with increasing the value of \textit{U}. The value of magnetic moment increases from -0.14 $\mu_{B}$ at \textit{U} = 1 eV to 0.55 $\mu_{B}$ at \textit{U} = 5 eV. There is also a very anomalous effect of \textit{U} on the magnetic moment of Fe atom, as the magnetic moment changes its sign after \textit{U} = 2 eV and coupled ferromagnetically with Mn atom. On the other hand the magnetic moment of Mn decreases from 2.24 $\mu_{B}$ at \textit{U} = 1 eV to 0.77 $\mu_{B}$ at \textit{U} = 5 eV. At \textit{U} = 2 eV, the magnetic moment of Fe atom become almost zero and the total magnetic moment is contributed only by the Mn atom. In the Fe atom the Hund's like exchange interactions are also found to increase with \textit{U} while in Mn atom decrease with \textit{U}. In comparison to Fe$_{2}$MnSi the effect of \textit{U} on the magnetic properties of Fe$_{2}$MnAl compound is found to be more.  

The on-site Coulomb interactions are affecting above two compounds drastically but we have observed no drastic effect on the magnetic moment of Co$_{2}$MnGe. In this compounds the number of electrons in spin-up and spin-dn channels are no changing significantly with \textit{U}. The magnetic moment of Co atom increases slightly from 1.07 $\mu_{B}$ at \textit{U} = 1 eV to 1.15 $\mu_{B}$ at \textit{U} = 5 eV. The magnetic moment of Mn atom decreases from 2.80 $\mu_{B}$ at \textit{U} = 1 eV to 2.61 $\mu_{B}$ at \textit{U} = 5 eV. 

The value of \textit{U} is affecting the magnetic properties of these compounds although it has very less effect on Co$_{2}$MnGe. Therefore we have also studied the electronic structures of all these compounds in the presence of \textit{U}. The total and partial DOS of Fe$_{2}$MnSi, Fe$_{2}$MnAl and Co$_{2}$MnGe are presented in Fig. 7-9 only for two selected values of \textit{U} i.e. \textit{U} = 2 and 4 eV .

The spin-up channel of Fig. 7(a) shows that the TDOS at \textit{E$_{F}$} decrease very slowly with \textit{U} and there exists a gap at \textit{E$_{F}$} in down spin channel. The value of band gap in spin-up channel increases from $\sim$0.8 eV at \textit{U} = 2 eV to $\sim$0.9 eV at \textit{U} = 4 eV. TDOS in the spin-up and spin-dn channels shifts towards the lower energy as the value of \textit{U} is increased from 2 eV to 4 eV. The PDOS of Fe atom is presented in Fig. 7(b) and it is evident from figure that for these two values \textit{U}, \textit{e$_{g}$} and \textit{t$_{2g}$} states are contributing to the TDOS at \textit{E$_{F}$} with more contribution from \textit{e$_{g}$} states. Also the contribution from \textit{t$_{2g}$} and \textit{e$_{g}$} states at \textit{E$_{F}$} decreases from $\sim$0.45 states/eV/atom at \textit{U} = 2 eV to $\sim$0.27 states/eV/atom at \textit{U} = 4 eV and $\sim$1.7 states/eV/atom at \textit{U} = 2 eV to $\sim$1.0 states/eV/atom at \textit{U} = 4 eV, respectively. The contribution from both these states decrease with \textit{U} and PDOS appears to shift towards lower energy in the both spin channels. The PDOS of Mn (Fig. 7(c)) at \textit{E$_{F}$} is occupied by \textit{t$_{2g}$} states with negligibly small contribution from \textit{e$_{g}$} states. Also the occupancy of \textit{t$_{2g}$} states decreases very slowly with \textit{U} and \textit{t$_{2g}$} and \textit{e$_{g}$} states shift towards lower energy in spin-up channel. In the down spin channel the \textit{t$_{2g}$} states also shift towards lower energy while, \textit{e$_{g}$} states shift towards higher energy. The PDOS of Si atom has negligibly small contribution from 3\textit{s} and 3\textit{p} orbitals for these values of \textit{U}.  

The TDOS of Fe$_{2}$MnAl for \textit{U} = 2 eV and 4 eV are shown in Fig. 8(a). For \textit{U} = 2 eV, the soft gap is appearing in the spin-up channel and when value of \textit{U} is increased this compound become semiconductor with band gap of $\sim$0.7 ev. In the spin-dn channel the band gap does not change significantly with \textit{U}. This is also observed in magnetic moment calculations, where at \textit{U} = 2 eV, the magnetic moment of Fe atom become almost zero and for \textit{U} = 4 eV the value of magnetic moment is positive. The TDOS for both the spin channels shift towards higher energy at \textit{U} = 4 eV. The PDOS of Fe, Mn and Al atoms shown in Fig. 8(b-d) for \textit{U} = 2 eV and \textit{U} = 4 eV. From PDOS of Fe atom it is clear that \textit{t$_{2g}$} and \textit{e$_{g}$} states shifts towards higher energy for both the spin channels. The PDOS of Mn atom show that in spin-up channel shifting of \textit{t$_{2g}$} states is dominating and in spin-dn channel shifting of \textit{e$_{g}$} states is more dominating. Also shifting is taking place at higher rate in the conduction band. The PDOS of Al atom has also negligibly small contribution from 3\textit{s} and 3\textit{p} orbitals for these values of \textit{U}. 

The TDOS of Co$_{2}$MnGe in the presence of \textit{U} are shown in Fig. 9(a). This figure shows that the spin-up channel of Co$_{2}$MnGe is metallic for all values of \textit{U} and down spin channel is semiconducting. As the value of \textit{U} is increased from 2 to 4 eV, there is broadening of the band gap in the spin-dn channel. The band gap is found to increase from  $\sim$1.0 eV at \textit{U} = 2 eV to  $\sim$1.4 eV at \textit{U} = 4 eV.  We have not observed any significant shifting in the TDOS for both spin channels. Thus the on-site Coulomb interactions are not playing a significant effect on this compound, which is also evident from the calculations of magnetic moments. It is clear from PDOS of Co atom shown in Fig. 9(b) that at \textit{E$_{F}$}, the \textit{t$_{2g}$} and \textit{e$_{g}$} states contribute same for both values of \textit{U}. Increasing value of \textit{U} is not going to affect the PDOS of Co atom. The PDOS of Mn atom (Fig. 9(c)) also shows no difference in the DOS at \textit{E$_{F}$} for these two values of \textit{U}. Fig. 9(d) shows that contribution from 3\textit{s} and 3\textit{p} states to the TDOS is negligibly small and also not affected by the on-site Coulomb interactions.  

From the above results it is clear that the metallic ground state of Fe$_{2}$MnAl compound changes directly to the semiconducting ground state when the value of \textit{U} is increased, whereas ground state of other two Heusler alloys remain half-metallic ferromagnetic. Very few experimental work related to Fe$_{2}$MnAl compound is found in the literature. In order to verify our predicted result it is necessary to perform electrical conductivity and neutron diffraction experiments, which directly probe the electronic transport behaviour and the nature of magnetic coupling between the Fe and Mn moments. To the best of our knowledge there are no experimental data on electrical conductivity in the $L2_{1}$ phase. However Liu et al. \cite{liu} have reported resistivity data of the compound when it is in the B$_{2}$ phase, which show insulating behaviour at the low temperature. Similarly neutron diffraction experiments on Fe$_{2}$MnSi compound are also desirable to know the magnetic moments of the Fe and Mn atoms. These experiments will help in understanding the role of on-site Coulomb interaction among the 3\textit{d} electrons of Fe and Mn by studying the magnitude and directions of the magnetic moments of Fe and Mn atoms in the Fe$_{2}$MnSi and Fe$_{2}$MnAl compounds.

\section{Conclusions}
The full-potential linearized augmented-plane wave methods have been employed to study the electronic and magnetic properties of Fe$_{2 }$MnSi, Fe$_{2 }$MnAl and Co$_{2 }$MnGe. The ferromagnetic (FM) solutions without using on-site Coulomb interaction \textit{U} show the presence of half-metallic FM ground state in Fe$_{2 }$MnSi however, in Fe$_{2 }$MnAl the ground state is found to be metallic. The total magnetic moment is contributed by Mn atom with small contribution from Fe atom in both cases. The electronic and magnetic properties of Fe$_{2 }$MnSi and Fe$_{2 }$MnAl are affected significantly by \textit{U}, whereas the almost negligible effect of \textit{U} is found in Co$_{2}$MnGe. The magnetic moment of Fe atom in Fe-based compounds is found to increase with \textit{U} and for Mn atom its value decreases. In Fe$_{2}$MnSi the Fe and Mn moments are coupled ferromagnetically for all values of U, whereas in Fe$_{2}$MnAl they coupled antiferromagnetically below U = 2 eV and ferromagnetically above it. The study of electronic structures show that in Fe$_{2 }$MnSi and Co$_{2}$MnGe the ground state remains half-metallic ferromagnetic for all values of \textit{U}, whereas in Fe$_{2 }$MnAl compound the ground state become semiconducting after \textit{U} = 2 eV. The ferromagnetic coupling between Fe and Mn moments appears to be responsible for this in Fe$_{2 }$MnAl compound.

\begin{table}
\begin{center}
\setlength{\tabcolsep}{5pt}
\caption{Calculated total number of electrons in spin-up (3\textit{d}$_{up}$) and spin-dn (3\textit{d}$_{dn}$) channels and partial magnetic moments m($\mu _{B}$) of Fe$_{2}$MnSi, Fe$_{2}$MnAl and Co$_{2}$MnGe with different values of \textit{U} } 
\centering
\begin{tabular}{ c c  c  c  c c c c}

 \hline\\
&&\multicolumn{3}{c}{Fe/Co atom}& \multicolumn{3}{c}{Mn atom} \\ [0.5ex]
Compound&\textit{U} (eV) & 3\textit{d}$_{up}$ & 3\textit{d}$_{dn}$ & m($\mu _{B}$) & 3\textit{d}$_{up}$ & 3\textit{d}$_{dn}$ & m($\mu _{B}$) \\ 
\hline\\ 
&1&3.21&2.92&0.29&3.67&1.26&2.41\\
&2&3.25&2.89&0.36&3.62&1.32&2.30\\
Fe$_{2}$MnSi&3&3.32&2.82&0.50&3.49&1.44&2.05\\
&4&3.39&2.74&0.65&3.34&1.59&1.75\\
&5&3.50&2.63&0.87&3.11&1.82&1.29\\ [1 ex]
&1&2.97&3.10&-0.14&3.55&1.31&2.24\\
&2&3.03&3.03&0.00&3.41&1.44&1.97\\
Fe$_{2}$MnAl&3&3.11&2.94&0.17&3.24&1.60&1.64\\
&4&3.20&2.84&0.36&3.03&1.81&1.22\\
&5&3.29&2.74&0.55&2.80&2.03&0.77\\ [1 ex]
&1&4.16&3.09&1.07&3.91&1.11&2.80\\
&2&4.16&3.08&1.08&3.89&1.13&2.76\\
Co$_{2}$MnGe&3&4.17&3.06&1.11&3.88&1.15&2.73\\
&4&4.17&3.05&1.12&3.85&1.17&2.68\\
&5&4.18&3.03&1.15&3.82&1.21&2.61\\ [1 ex]
\hline
\end{tabular}
\end{center}
\end{table}

\begin{figure}
    \centering
    \includegraphics[width=0.8\textwidth,natwidth=610,natheight=642]{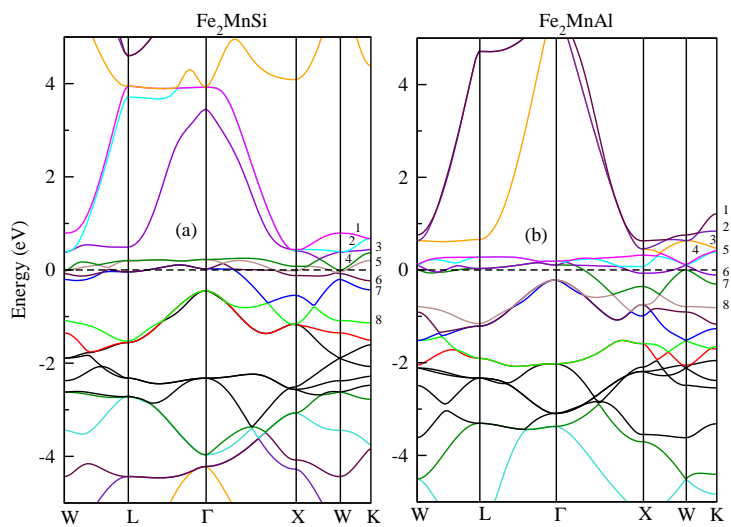}
\caption{(Color online) Electronic band structures in the paramagnetic phase for 
 (a) Fe$_{2}$MnSi and (b) Fe$_{2}$MnAl.}
\label{Fig1}
\end{figure}

\begin{figure}
    \centering
    \includegraphics[width=0.8\textwidth,natwidth=610,natheight=642]{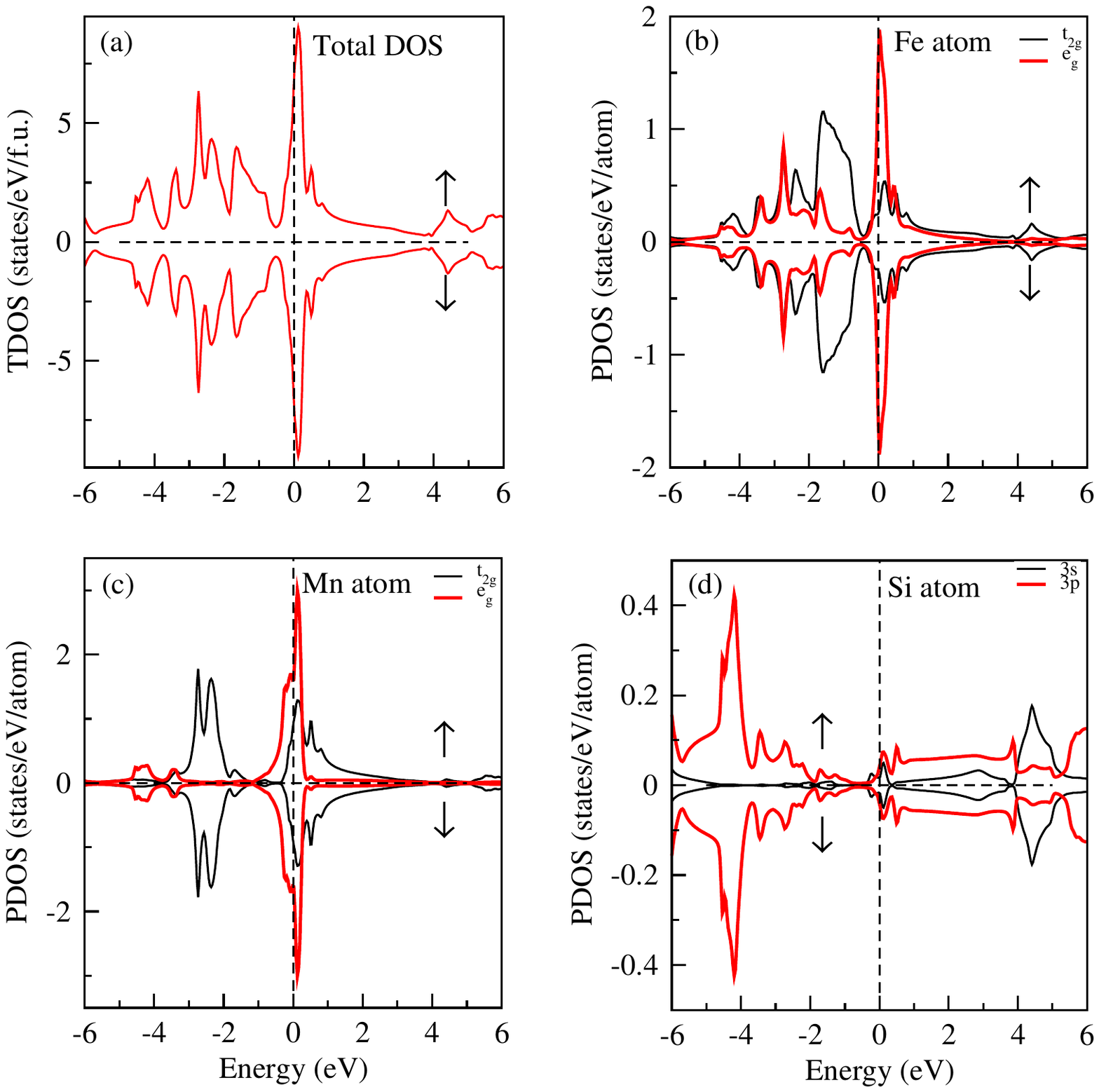}
\caption{(Color online) Total and partial density of states plots for
Fe$_{2}$MnSi in the paramagnetic phase. Shown are (a) the TDOS plot,
(b) PDOS of Fe atom, (c) PDOS of Mn atom and (d) PDOS of Si
atom.}
\label{Fig2}
\end{figure}

\begin{figure}
    \centering
    \includegraphics[width=0.8\textwidth,natwidth=610,natheight=642]{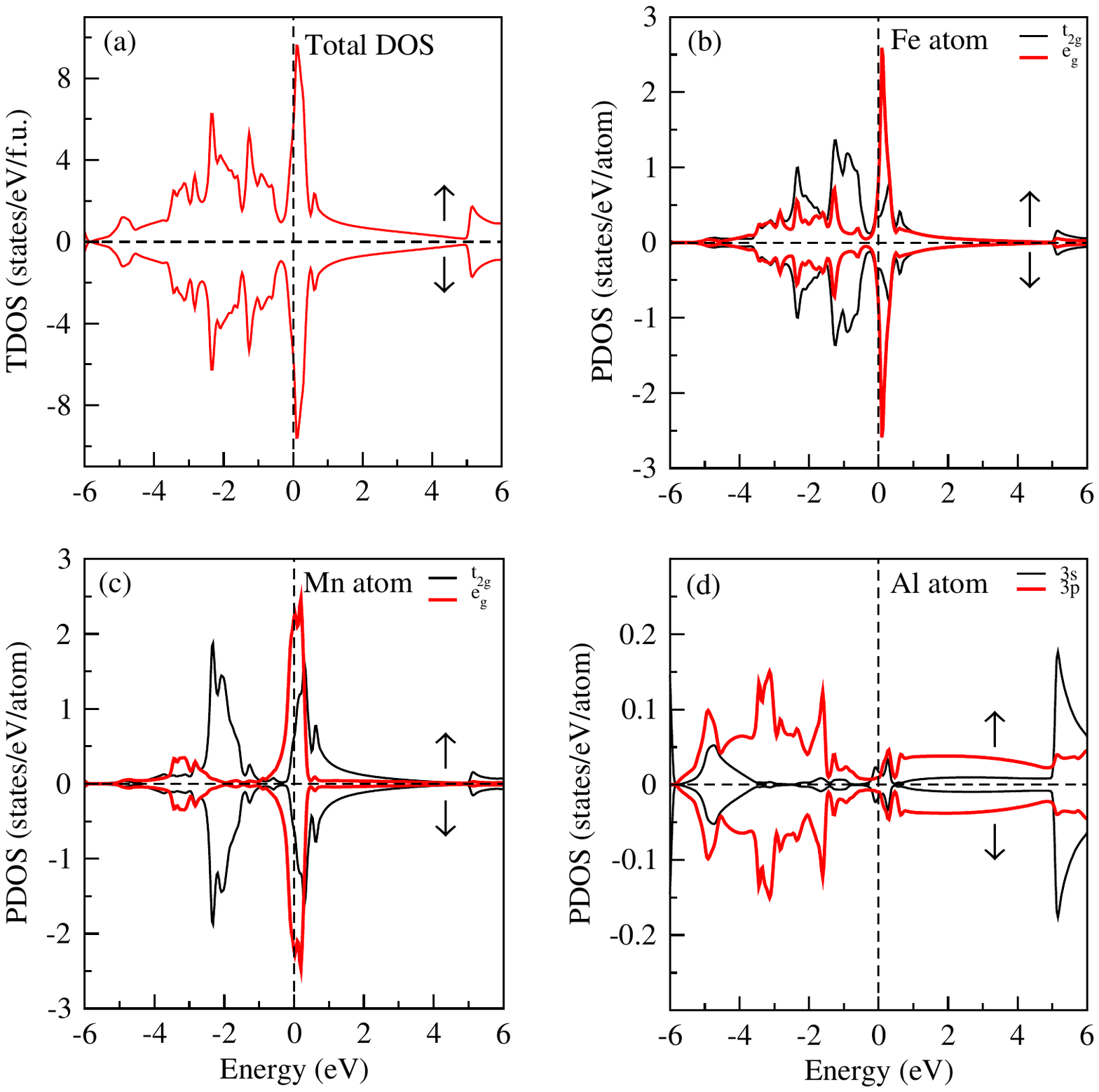}
\caption{(Color online) Total and partial density of states plots for
Fe$_{2}$MnAl in the paramagnetic phase. Shown are (a) the TDOS plot,
(b) PDOS of Fe atom, (c) PDOS of Mn atom and (d) PDOS of Al
atom.}
\label{Fig3}
\end{figure}

\begin{figure}
    \centering
    \includegraphics[width=0.8\textwidth,natwidth=610,natheight=642]{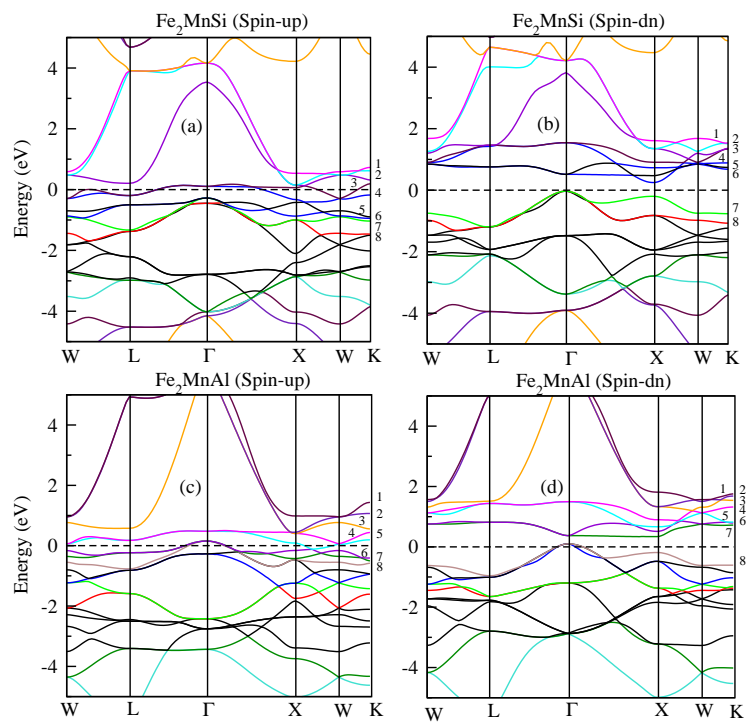}
\caption{(Color online) Electronic band structures in the ferromagnetic phase (a and b) for spin-up and spin-dn channels of Fe$_{2}$MnSi and (c and d) for spin-up and spin-dn channels of Fe$_{2}$MnAl.}
\label{Fig4}
\end{figure}

\begin{figure}
    \centering
    \includegraphics[width=0.8\textwidth,natwidth=610,natheight=642]{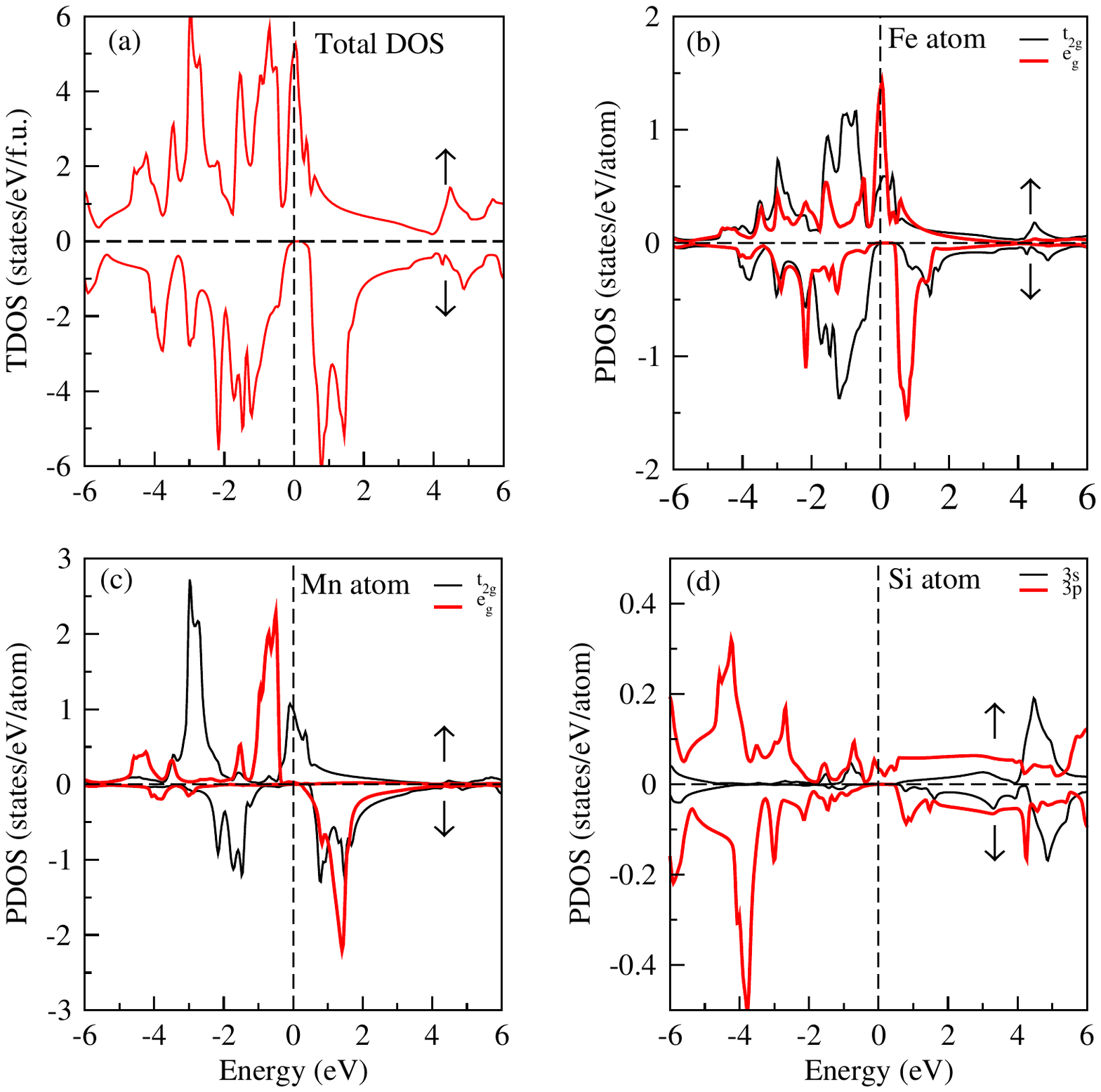}
\caption{(Color online) Total and partial density of states plots for
Fe$_{2}$MnSi in the ferromagnetic phase. Shown are (a) the TDOS plot,
(b) PDOS of Fe atom, (c) PDOS of Mn atom and (d) PDOS of Si
atom.}
\label{Fig5}
\end{figure}

\begin{figure}
    \centering
    \includegraphics[width=0.8\textwidth,natwidth=610,natheight=642]{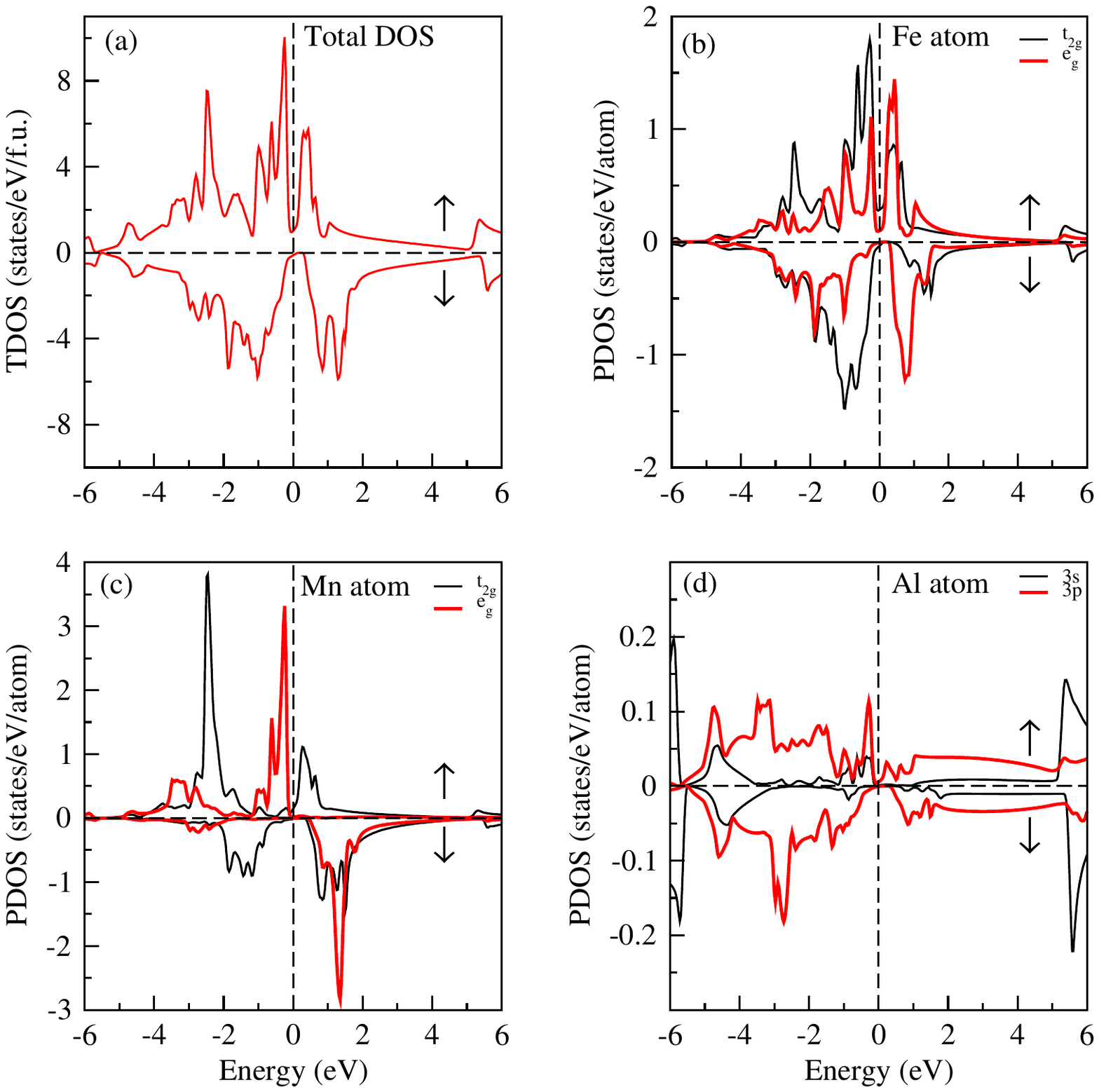}
\caption{(Color online) Total and partial density of states plots for
Fe$_{2}$MnAl in the ferromagnetic phase. Shown are (a) the TDOS plot,
(b) PDOS of Fe atom, (c) PDOS of Mn atom and (d) PDOS of Al
atom.}
\label{Fig6}
\end{figure}

\begin{figure}
    \centering
    \includegraphics[width=0.8\textwidth,natwidth=610,natheight=642]{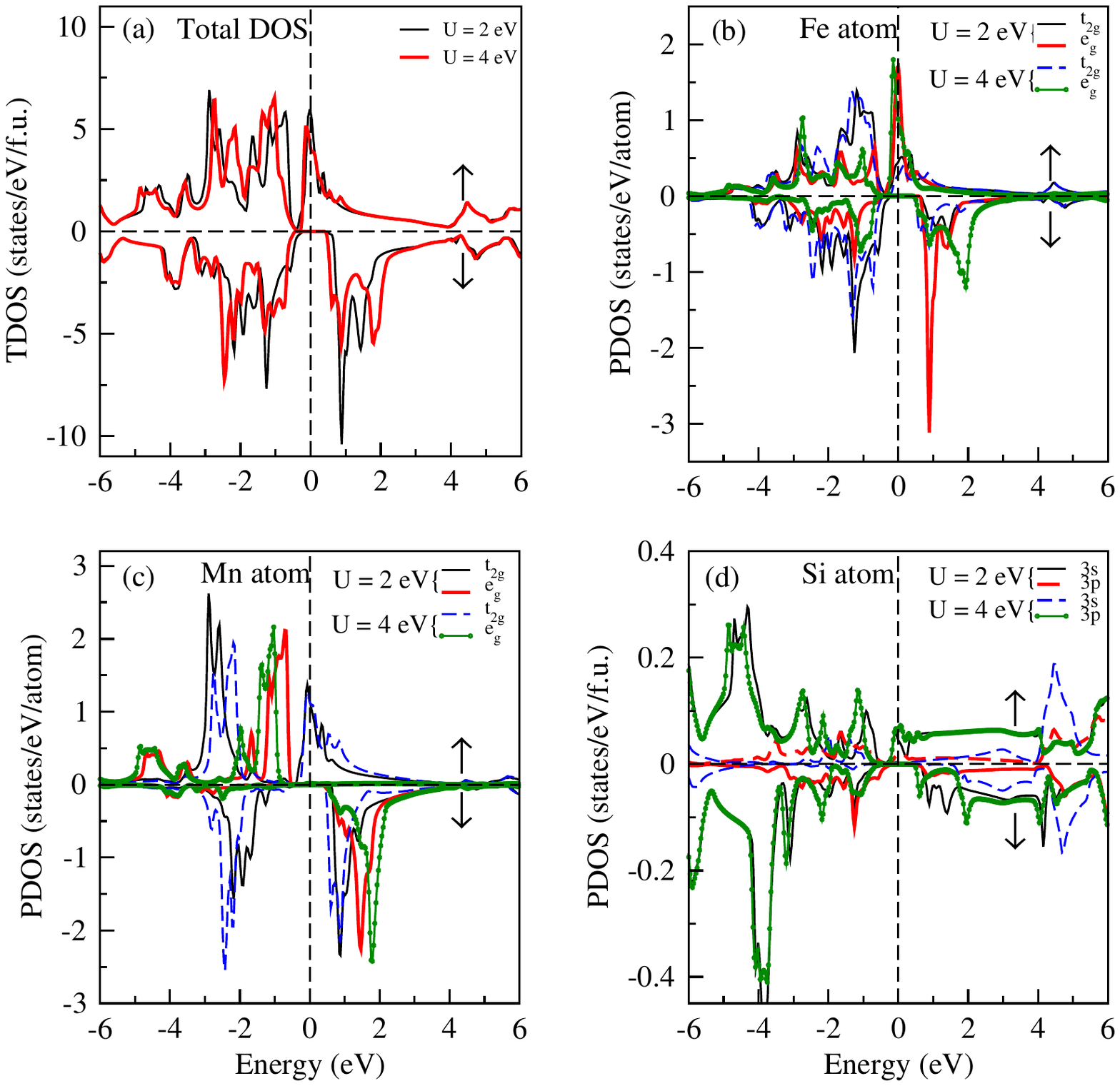}
\caption{(Color online) Total and partial density of states plots for
Fe$_{2}$MnSi in the ferromagnetic phase in the presence of \textit{U}. Shown are (a) the TDOS plot,
(b) PDOS of Fe atom, (c) PDOS of Mn atom and (d) PDOS of Si
atom.}
\label{Fig7}
\end{figure}

\begin{figure}
    \centering
    \includegraphics[width=0.8\textwidth,natwidth=610,natheight=642]{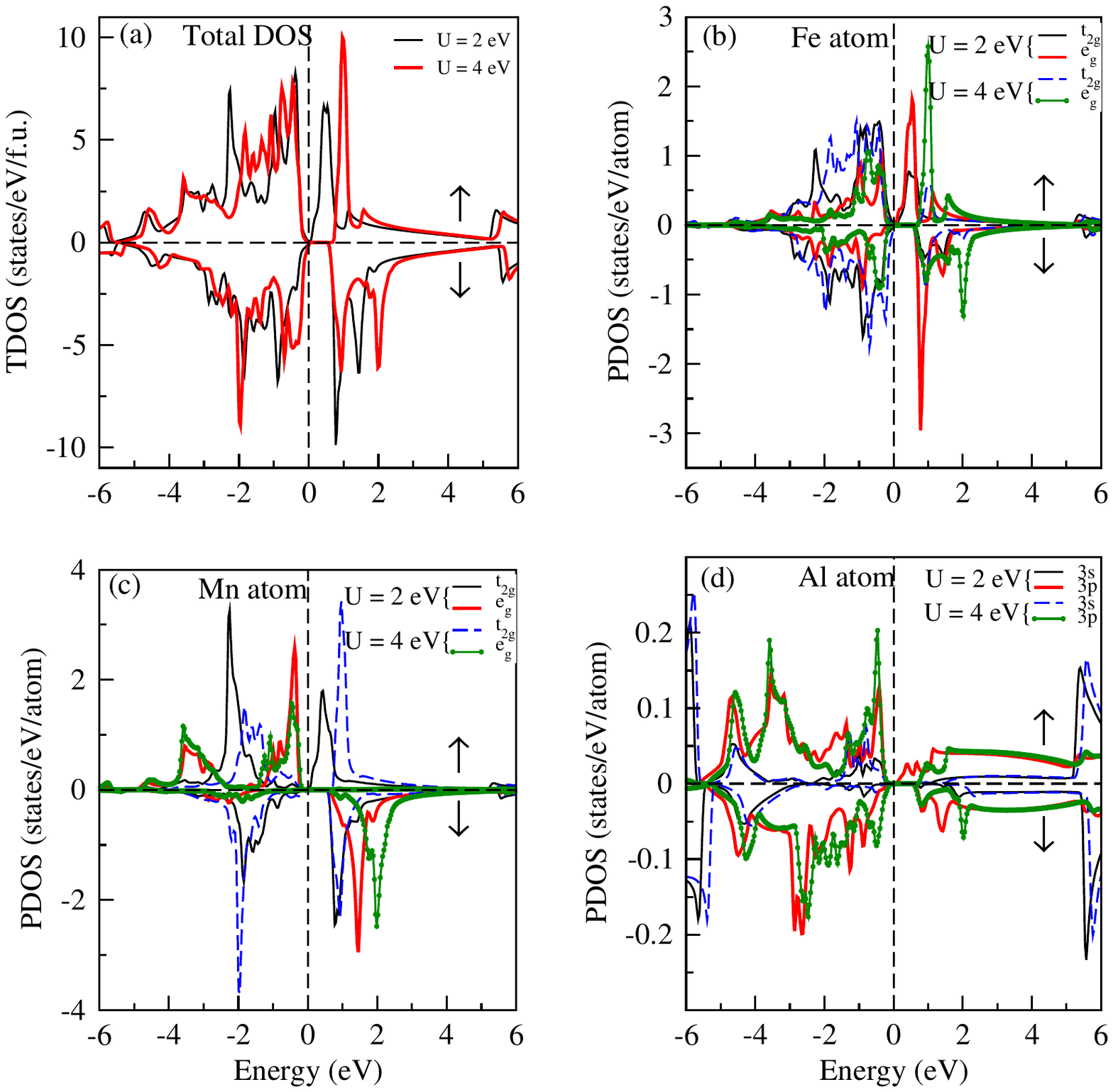}
\caption{(Color online) Total and partial density of states plots for
Fe$_{2}$MnAl in the ferromagnetic phase in the presence of \textit{U}. Shown are (a) the TDOS plot,
(b) PDOS of Fe atom, (c) PDOS of Mn atom and (d) PDOS of Al
atom.}
\label{Fig8}
\end{figure}

\begin{figure}
    \centering
    \includegraphics[width=0.8\textwidth,natwidth=610,natheight=642]{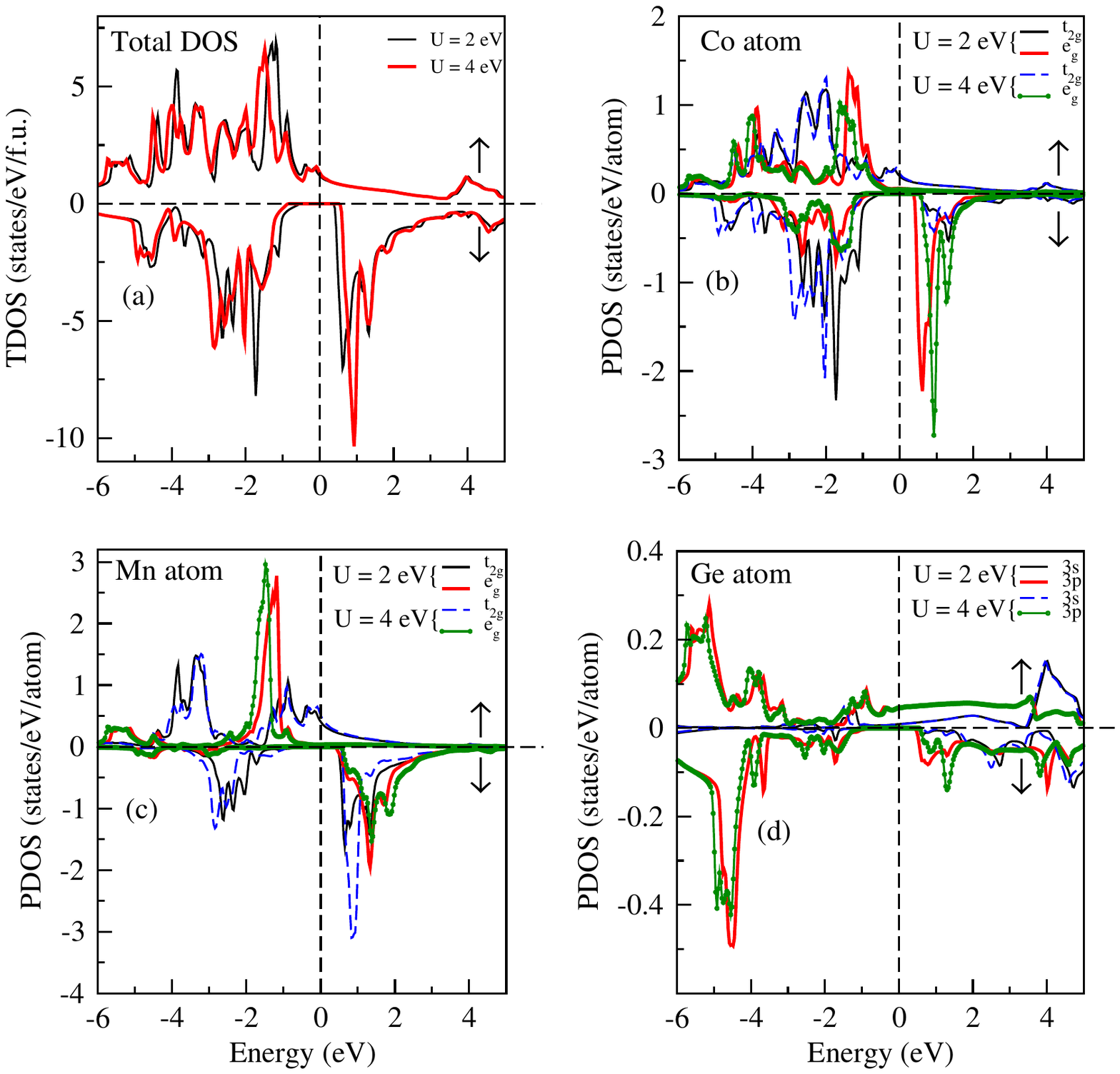}
\caption{(Color online) Total and partial density of states plots for
Co$_{2}$MnGe in the ferromagnetic phase in the presence of \textit{U}. Shown are (a) the TDOS plot,
(b) PDOS of Co atom, (c) PDOS of Mn atom and (d) PDOS of Ge
atom.}
\label{Fig9}
\end{figure}

\end{document}